\documentclass[conference]{IEEEtran}
\IEEEoverridecommandlockouts
\usepackage{cite}
\usepackage{xurl}
\usepackage{float}
\usepackage{amsmath,amssymb,amsfonts}
\usepackage{algorithmic}
\usepackage{graphicx}
\usepackage{minted}
\usepackage{textcomp}
\usepackage{xcolor}
\usepackage{enumitem}
\usepackage[
  colorlinks=true,
  linkcolor=blue,
  citecolor=blue,
  urlcolor=blue
]{hyperref}
\usepackage{url}
\usepackage{titlesec}
\renewcommand{\thesubsection}{\Alph{subsection}}
\makeatletter
\def\@IEEEsectpunct{}
\makeatother

\def\BibTeX{{\rm B\kern-.05em{\sc i\kern-.025em b}\kern-.08em
    T\kern-.1667em\lower.7ex\hbox{E}\kern-.125emX}}

\begin{document}

\title{TessPay: Verify-then-Pay Infrastructure for Trusted Agentic Commerce}

\author{
    \IEEEauthorblockN{
        Mehul Goenka\textsuperscript{1,3}, 
        Tejas Pathak\textsuperscript{2,3}, 
        Siddharth Asthana\textsuperscript{1,3}
        \\[0.2em] 
        \normalfont\small 
        \textsuperscript{}                    \\
        \textsuperscript{1}University of Oxford\\
        \textsuperscript{2}Indian Institute of Technology, Delhi\\
        \textsuperscript{3}Tesseris.org
    }
}

\maketitle

\begin{abstract}
The global economy is entering the era of Agentic Commerce, where autonomous agents can discover services, negotiate prices, and transact value. However adoption towards agentic commerce faces a foundational trust gap: current systems are built for direct human interactions rather than agent-driven operations. It lacks core primitives across three critical stages of agentic transactions. First, Task Delegation lacks means to translate user intent into defined scopes, discover appropriate agents, and securely authorize actions. Second, Payment Settlement for tasks is processed before execution, lacking verifiable evidence to validate the agent's work. Third, Audit Mechanisms fail to capture the full transaction lifecycle, preventing clear accountability for disputes. While emerging standards address fragments of this trust gap, there still remains a critical need for a unified infrastructure that binds the entire transaction lifecycle.

To resolve this gap, we introduce TessPay, a unified infrastructure that replaces implicit trust with a 'Verify-then-Pay' architecture. It is a two plane architecture separating control and verification from settlement. TessPay operationalizes trust across four distinct stages: Before execution, agents are anchored in a canonical registry and user intent is captured as verifiable mandates, enabling stakeholder accountability. During execution, funds are locked in escrow while the agent executes the task and generates cryptographic evidence (TLS Notary, TEE etc.) to support Proof of Task Execution (PoTE). At settlement, the system verifies this evidence and releases funds only when the PoTE satisfies verification predicates; modular rail adapters ensure this PoTE-gated escrow remains chain-agnostic across heterogeneous payment rails. After settlement, TessPay preserves a tamper-evident audit trail to enable clear accountability for dispute resolution.
\end{abstract}


\section{Introduction}
A new era of the digital economy is emerging as AI agents evolve from interactive, user-guided assistants to autonomous economic actors. These agents can increasingly interpret high-level intent, decompose it into multi-step workflows, and execute real actions across tools and services including payment settlement. This shift is reshaping how value is created and exchanged, giving rise to the era of agentic commerce. In this model, commerce moves from human-operated clicks to autonomous agent operation: agents operate the entire transaction lifecycle from discovering services, negotiating value, to settling payments on behalf of users\cite{ref2}.

This represents an evolution akin to the e-commerce revolution, where volumes of transactions increasingly originate from AI agent-mediated decision loops rather than direct human interaction \cite{ref1}. This "intent-to-settlement" economy spans all the way from travel booking and subscriptions to enterprise procurement and financial treasury operations. As these workflows become multi-agent by default, the core challenge shifts from agent capability to whether users can safely delegate authority and verify what happened when an autonomous agent moves real money.

However, the shift toward agentic commerce remains constrained by a fundamental trust gap. Today’s commerce infrastructure is built for direct human interaction instead of agent-driven operation. It lacks core primitives across three stages of agentic transactions:
\begin{enumerate}
    \item \textit{Task Delegation}: While agentic commerce requires delegating authority to non-human actors, existing infrastructure lacks mechanisms to  translate user intent into defined scopes, discover appropriate agents, and securely authorize actions throughout the transaction cycle—from intent to final settlement.

    \item \textit{Verified Settlement}: Even when task delegation aligns with user intent, payment of services is done before the agent executes the task, relying on the assumption that the agent will perform the task correctly. Therefore, it lacks verifiable evidence that the execution strictly adhered to the user's intent and the task scope.

    \item \textit{Dispute Resolution}: When things go wrong, the lack of an effective audit mechanism throughout the transaction lifecycle prevents clear accountability. Without this transparency, resolving disputes and processing refunds or chargebacks becomes operationally complex as it lacks a verifiable basis for mediation.

\end{enumerate}
While promising progress has emerged through new protocols \cite{ref3} and research, today’s landscape addresses the trust gap only in fragments. Google’s AP2 protocol uses Verifiable Digital Credentials (VDCs)\cite{ref12} to formalize intent and delegated authority, but it functions primarily as a coordination layer and does not address verification before settlement. Coinbase’s x402 \cite{ref7} enables pay-per-request payments, but it follows a “pay-first, then response” model, which is misaligned with high-value transactions that require outcome verification prior to releasing funds. ERC-8004\cite{ref10} strengthens agent identity and discoverability through a universal registry, yet it does not address the trust gap of verified settlement. In parallel, research efforts such as Agent-PROV\cite{ref26} and WebProofs\cite{ref27} advance provenance and cryptographic evidence for off-chain actions, but are not integrated with the rest of the stages of lifecycle including authorisation, delegation and settlement. Consequently, there remains a need for an infrastructure that unifies delegation, secure execution, verifiable outcomes, and fair settlement into a single auditable lifecycle. 

Bridging these gaps requires infrastructure that treats delegation, execution, and settlement as a single verifiable lifecycle. At the control layer, agents must possess stable identities anchored in a canonical registry to prevent impersonation and bind actions in immutability. This layer must further support efficient discovery and multi-agent orchestration, allowing dynamic workflow composition without losing provenance of authority. At the verification and settlement layer, value transfer must be conditional: funds are locked under user-approved terms and released only when execution artifacts provide cryptographically substantiated proof, backed by end-to-end audit trails for accountability and dispute resolution.

To address these gaps, we introduce TessPay, a unified infrastructure utilising a "Verify-then-Pay" model that couples task delegation, verified settlement, and dispute resolution into a single transaction lifecycle. TessPay addresses this gap of trust across four different stages: Before execution, agents are anchored in a stable identity via a canonical registry, and the user's intent is captured as verifiable mandates enabling clear stakeholder accountability and intent-driven service discovery.  During execution, funds are locked in escrow while the agent executes the task and generates cryptographic evidence (TLS Notary, TEE etc.) to support Proof of Task Execution (PoTE). At settlement, the system verifies this evidence and releases funds only when the PoTE satisfies explicit verification predicates. The settlement is chain-agnostic enabled by modular payment rail integration. After settlement, TessPay preserves a tamper-evident audit trail to enable dispute resolution, refunds, and clear accountability whenever task execution deviates from the authorized scope.

\subsection*{\textbf{Contributions}}
TessPay makes the following key contributions:
\begin{itemize}
    \item An architecture that unifies agent identity, service discovery, task delegation, output verification, and payment settlement into a modular microservices infrastructure

    \item A "Verify-then-Pay" settlement model that holds funds in escrow and releases it upon receiving a proof of task execution (PoTE) supported by verifiable cryptographic evidence (TLS Notary, TEE etc.)

    \item Chain-agnostic settlement model that supports settlement across heterogeneous chains enabled by modular integration of payment rail adapters

    \item A tamper-resistant audit trail that captures verifiable traces across the payment lifecycle providing clear accountability in case of disputes, refunds and chargebacks

\end{itemize}

In this paper, we first review related work in the field of agentic commerce and identify the fragmentation and unresolved gaps between current approaches. Using this, we then derive a set of design goals that directly address these gaps and act as guiding principles for our architecture.

Next, we introduce TessPay as a two-plane architecture that separates control and verification from settlement. We explore the core components and microservices that build up the architecture, and detail the internal process flows that operationalize intent capture, delegation, verification, and conditional settlement across the payment lifecycle. We then discuss TessPay’s integration across heterogeneous chains and standards with its modular payment rail design. We then detail the three transaction tiers it supports and the verification mechanisms associated with each tier, and illustrate a complete end-to-end payment flow.

Finally, we evaluate TessPay’s security properties through threat modeling and describe the mechanisms used to mitigate key vulnerabilities. We then present representative use cases of an e-commerce agent and a portfolio manager agent to demonstrate how TessPay resolves the infrastructural bottleneck and enables verified agentic payments. We conclude by outlining directions for future work and summarizing the paper’s contributions.

\section{Background and Related Work}
Current developments in agentic commerce can be categorized into protocol specifications and theoretical frameworks addressing the economic risks of autonomy. Although there are now foundational protocols for agents to negotiate terms and initiate payments, the frameworks for securely settling these transactions are still evolving. This section reviews these strands of work, contrasting the emerging industrial standards for agentic payment with academic research that underscores the barriers to widespread financial delegation.

On the industrial front, Google’s Agent Payments Protocol (AP2) \cite{ref5} has established a robust standard for the ``language of intent,'' introducing signed Verifiable Digital Credentials to capture user intent and trust delegation  before a transaction occurs. However, AP2 functions strictly as a messaging layer, it relies on traditional payment rails and assumes merchant honesty, lacking an intrinsic mechanism to verify execution before funds are captured.

Simultaneously, the x402 protocol\cite{ref8}, pioneered by Coinbase, revives the HTTP 402 status code\cite{ref9} to enable ``pay-per-request'' interactions. While efficient for low-stakes API gating, x402 enforces an optimistic ``pay-then-consume'' logic where funds are released prior to service delivery, exposing the principal to the risk of non-delivery or hallucinated outputs.

Similarly, the Agent Commerce Protocol (ACP) \cite{ref11} structures the negotiation lifecycle but relies on reputation scoring for evaluation. This creates a ``cold start'' vulnerability where new agents cannot prove their reliability without a prior transaction history, necessitating a shift from reputation-based trust to deterministic proofs of execution. Even with the emergence of specialized protocols like Google’s AP2, Coinbase’s x402, and ERC-8004, they still remain isolated, forcing a ``fragmentation'' in the industry and preventing a unified agentic payment flow where these technologies can function together.

This lack of integration is compounded by a profound trust gap, where the absence of pre-settlement verification and transparent audit rails makes autonomous transactions inherently risky, as there is no cryptographic proof that a task was completed as intended before value is transferred. Furthermore, the ecosystem suffers from severe interoperability friction, where agents are often locked into proprietary platforms (like Google or specific L1 chains), hindering the realization of a truly chain-agnostic and discoverable agentic economy. Most importantly, the coupling of control and settlement where agents are granted access to private keys---creates an unacceptable attack surface for prompt injections and hallucinations leading to a lot of capital loss.

TessPay seeks to resolve these challenges by providing a universal, verified infrastructure that strictly decouples AI reasoning from financial execution, ensuring that autonomous commerce is as secure and interoperable as it is efficient.

\section{Design Goals}

\subsection{Unification of Agentic Commerce Primitives}
With so many recent developments in the field of agentic commerce with the incoming of protocols and solutions like Google's A2A \cite{ref4}, AP2\cite{ref6}, Coinbase's x402 , ERC-8004 for indexing and TEE Attestation for verification, there is still a problem---while they address pieces of the infrastructural trust gap, they are yet fragmented. There does not exist a unified single flow where all of these protocols are used in tandem with each other. One of the key design goals here is to create a unified agentic payment infrastructure that integrates all of these developments in a single seamless flow.

\subsection{Enabling Trust through Transparency}
A central challenge in agentic commerce is trust: even when an agent can initiate and execute transactions autonomously, a fundamental risk remains---whether the promised task was completed as agreed. This necessitates a verified agentic commerce model in which task execution is validated prior to the settlement based on the user intent and defined scope. In addition, the system must provide end-to-end auditability, maintaining a tamper-evident audit trail at each step of the workflow to enable clear accountability for dispute resolution. 

\subsection{Ensuring Interoperability across Ecosystems}
As agentic commerce scales, interoperability becomes a primary bottleneck: an agent operating within the Google ecosystem cannot seamlessly transact with a decentralized service on Base without significant friction across discovery and settlement. Accordingly, a key design goal is to serve as a universal infrastructure layer that bridges commerce protocols, indexing services, and settlement networks---ensuring that agents operating via TessPay remain discoverable to external registries, while enabling chain-agnostic settlement so payments can be executed on any supported integrated chain.

\subsection{Separation of Control and Settlement Planes}
A critical failure mode in many existing agentic payment systems is that agents are granted direct access to a user’s private keys or a master account, creating an unacceptable attack surface in which a single prompt injection or model hallucination can trigger unauthorized transfers and catastrophic fund loss. Realizing this, another one of our key design goals here is to enforce a strict separation of concerns between the control plane and the settlement plane, exposing only a minimal, well-defined interface of essential functions between the two \cite{ref23}.

\section{Architecture}

\begin{figure*}[htbp]
    \centering
    \includegraphics[width=\textwidth]{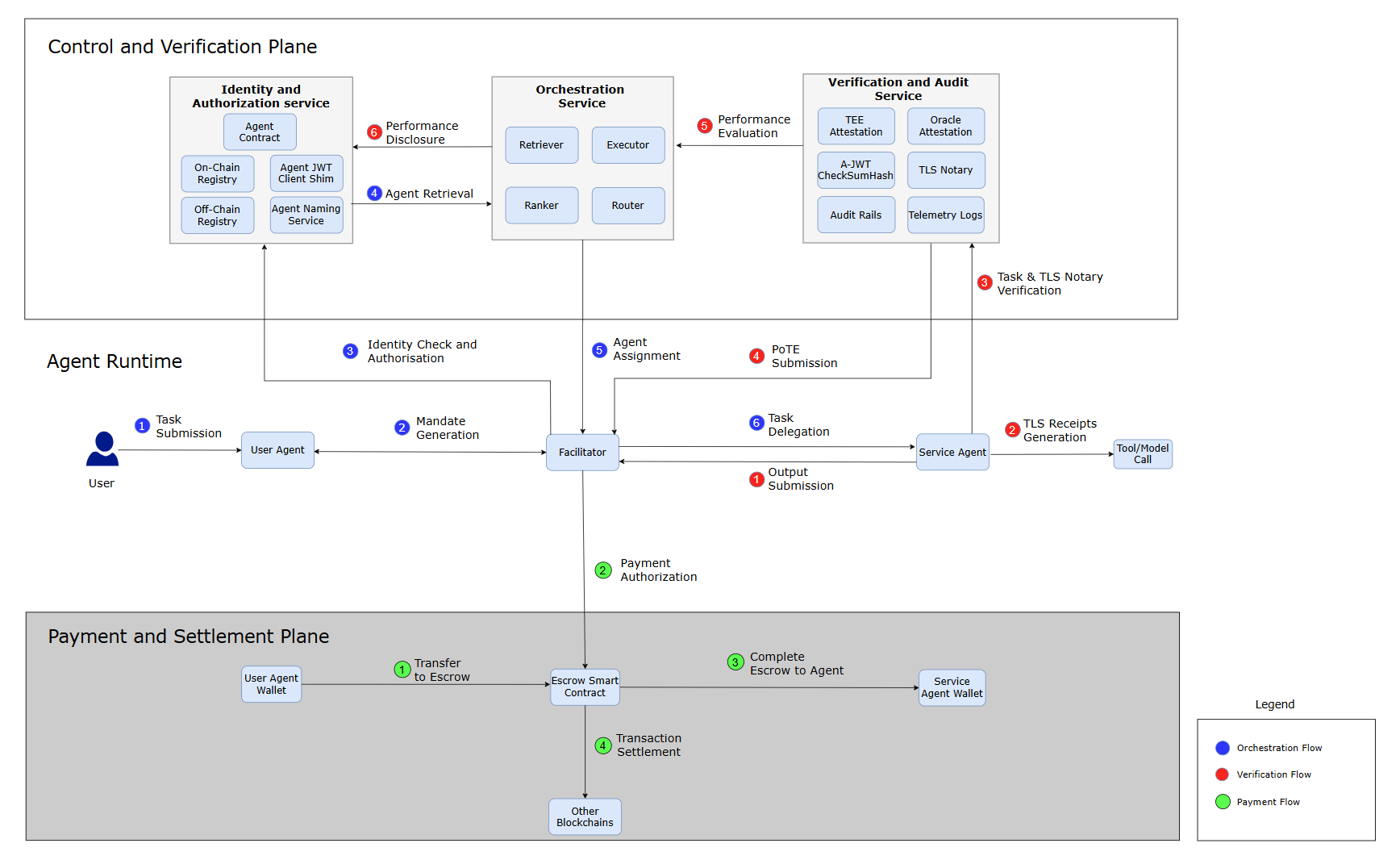} 
    \caption{\textbf{TessPay Internal Architecture :} TessPay separates coordination from value movement: the Control and Verification Plane (identity/authorization, orchestration, verification/audit services) governs mandates, A-JWT issuance, PoTE validation, telemetry, and audit rails; the Agent Runtime (user agent → facilitator → worker agent) executes tasks and produces receipts; and the Payment and Settlement Plane uses an escrow smart contract and rail adapters to lock and settle funds across external chains. Green/blue/red arrows denote payment, orchestration, and verification flows respectively.}
    \label{fig1}
\end{figure*}

To satisfy the above mentioned design goals we propose the following architecture for the TessPay payment Infrastructure. We propose the architecture as a Two-Plane Model, wherein the infrastructure is decomposed into a set of discrete, loosely coupled microservices. This architectural paradigm enforces a strict separation of concerns, distinctly isolating the domains of control and settlement. By decoupling these critical functions, the system establishes a resilient framework capable of mitigating the risks associated with the stochastic and non-deterministic nature of autonomous agent workflows.

To understand the TessPay architecture we explore it in the following progression :-
\begin{enumerate}[label=\Alph*.]
    \item Define the Entities in Agent Runtime
    \item Define the Two-Plane Architecture
    \item Explore the Process Flows
\end{enumerate}
\vspace{\baselineskip}

\subsection{Define the Entities in Agent Runtime}

\begin{itemize}[label=$\bullet$]
    \item \textbf{User Agent:} The User Agent (see \textit{Agent Runtime in Fig.~1}) is the primary interaction point for the end-user in TessPay. It orchestrates the negotiation phase by interacting with Merchant Agents through the AP2 protocol to discover offers, finalize quotes, and determine payment requirements. Crucially, the User Agent governs the collaborative construction of Intent, Cart, and Payment Mandates \cite{ref13, ref14}, ensuring that all encoded parameters---specifically budgetary limits, settlement rail selection, and temporal validity windows---rigorously adhere to the user’s defined preferences. Furthermore, it acquires an Agentic JWT token from the Authorization Service generated by explicitly capturing user intent and scoped strictly to the specific workflow step and for signing the Payment Mandate. This token is also used for accessing user's sensitive credentials by authorization based on user intent.

    \item \textbf{Facilitator:} The Facilitator functions as the central coordination unit within the Orchestration Service, responsible for resolving high-level user intent into executable multi-agent workflows. The facilitator orchestrates the entire payment lifecycle from user intent to settlement. It dynamically computes pricing, assesses risk parameters, and selects optimal settlement rails to construct the cryptographic Cart Mandate, thereby formalizing the specific commercial items and terms of the transaction. It manages task delegation to service agents while maintaining a verifiable trace of delegation. Furthermore, the Facilitator synchronizes the Settlement Plane with the execution in Control Plane by emitting critical lifecycle events to prime the Proof of Task Execution (PoTE) pipeline and trigger state transitions within the Payment Service.

    \item \textbf{Service Agent:} The Service Agent functions as the discrete execution unit of the multi-agent workflow responsible for fulfilling tasks delegated by the Facilitator. It interacts with computational models and external services strictly within the authorization boundaries defined by the Agentic JWT (A-JWT) and upstream Mandates by capturing user-intent. During operation, the agent emits verifiable execution traces, comprising of user- intent authorisation, model call, tool invocations, and other output proof artifacts  which are ingested by the Verification Service to construct the Proof of Task Execution (PoTE). Furthermore, the agent exposes granular runtime telemetry, including latency, token consumption, and computational costs, to the Telemetry Service to enable quantitative performance auditing and Service Level Agreement (SLA) enforcement.
\end{itemize}

\vspace{\baselineskip}

\subsection{Define the Two-Plane Architecture}

\subsubsection{Control and Verification Plane} 
This plane supports the payment settlement layer by serving as the primary layer for co-ordination, control and verification. This plane is visualized to be made up of these key microservices (See \textit{Fig.~1 Control and Verification Plane}):

\begin{enumerate}[label=\roman*.]
    \item \textbf{Identity and Authorization Service:} This service (see \textit{Fig.~1}) is responsible for providing authorization of user intent and providing discovery layer with access to agent identity and metadata with the Agent Registry. This service is made up of five components:

    \begin{itemize}[label=$\bullet$]
        \item \textbf{Agent Contract:} The Agent Contract is a smart contract that specifies the proof artifacts, verification requirements, and validation logic for successful verification of task execution by the agent. Concretely, it defines a verification predicate that commits to a fixed set of required proof objects and the corresponding validation mechanisms that must be satisfied for verification to succeed. For example, an order receipt in a e-commerce workflow.
        \item \textbf{Off-Chain Registry:} The off-chain registry serves as the canonical database of agents in the ecosystem, storing each agent’s identity attributes and associated metadata. These records are represented as a structured agent manifest (off-chain agent card) with sufficient detail to support interoperability, enabling agents to be indexed and discovered across heterogeneous registry and discovery services.
        \item \textbf{On-Chain Registry:} The on-chain registry anchors an agent’s identity, ownership, and the integrity of its off-chain manifest in an immutable blockchain record. Concretely, it binds the agent identifier to an ownership reference and stores a cryptographic commitment to the off-chain manifest, thereby grounding the manifest in on-chain immutability and enabling tamper-evident verification of the referenced metadata.
        \item \textbf{Agent Naming Service:} The Agent Naming Service generates human-readable agent domains analogous to Web2 naming conventions and assigns a unique \texttt{agent\_id} that serves as the canonical identity anchor. This identifier enables consistent resolution and recognition of agents across the distributed ecosystem.
        \item \textbf{Agent JWT Client Shim:} The Agent JWT Client handles the authorisation of user intent by issuing an Agent JWT token~\cite{ref25}. It computes a real-time cryptographic sum of the agent’s identity by hashing its defining characteristics (like system prompt and tool configurations) and embeds this identity into every authorization request. The shim tracks agent state and generates ephemeral Proof-of-Possession (PoP) keys to prevent replay attacks
    \end{itemize}

\begin{figure}[htbp]
    \centering
    \includegraphics[width=0.9\linewidth, keepaspectratio]{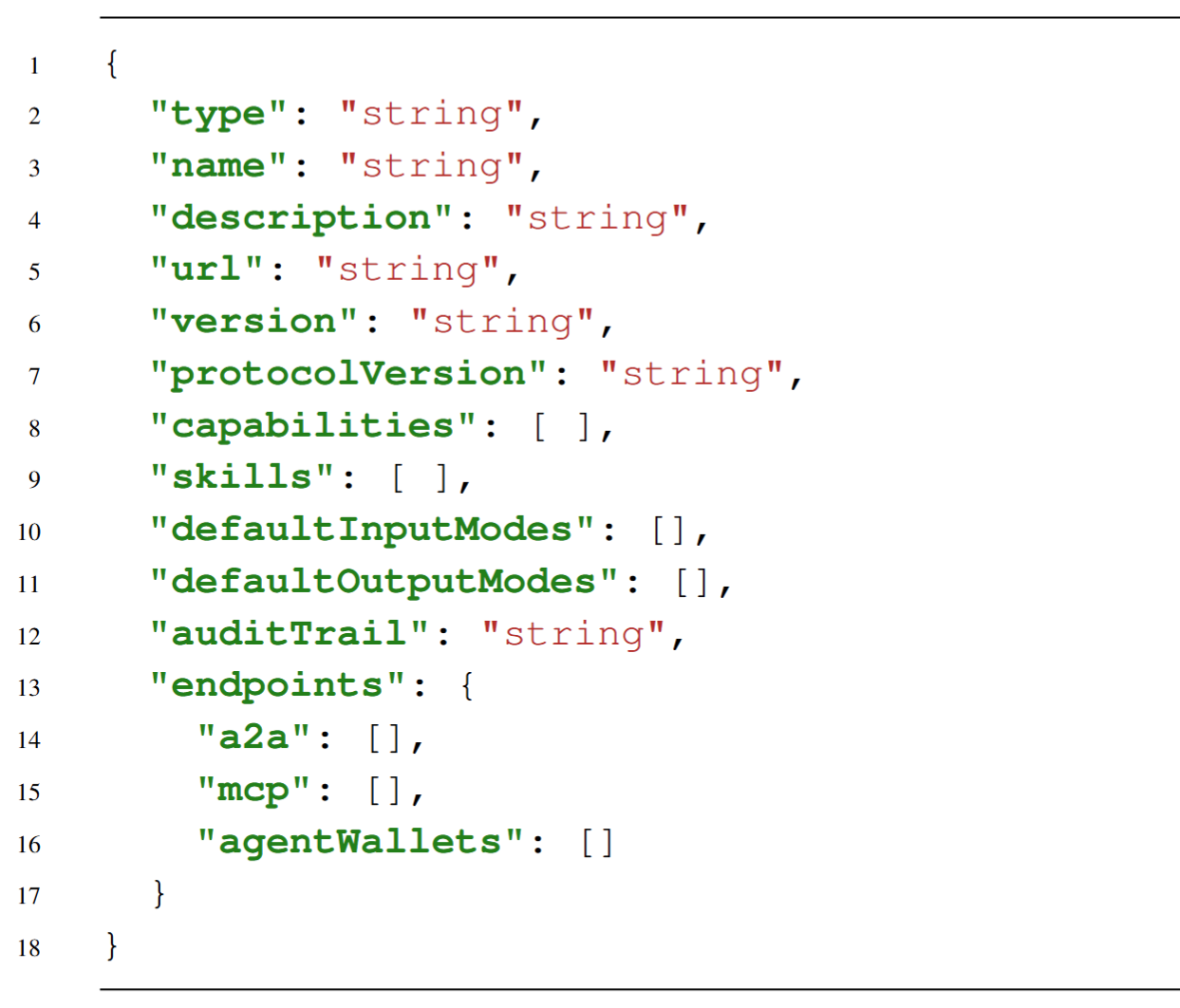}
    \caption{An Example Agent Manifest}
    \label{fig:json_schema}
\end{figure}
  
    \item \textbf{Orchestration Service:} The orchestration service (see \textit{Fig.~1}) is responsible for agent discovery and co-ordination of the flow across user intent and payment settlement. The orchestration is built up of the following key components:
    \begin{itemize}
        \item \textbf{Retriever:} It is responsible for querying the registry for the agent based on the required capability and returning a set of agents.
        \item \textbf{Ranker:} It is responsible for ranking the agents returned by the retriever for the given task.
        \item \textbf{Router:} It is responsible for making the routing decision between the top two agents returned by ranker.
        \item \textbf{Executor:} It is responsible for delegation of tasks to the service agent and also captures telemetry data.
    \end{itemize}

    \item \textbf{Verification and Audit Service:} This service verifies an agent’s task execution by validating the associated Proof of Task Execution (PoTE) and anchors the end-to-end payment flow in tamper-evident audit trails to ensure future auditability (see \textit{Fig.~2}). This service consists of the following components:
    
    \begin{itemize}
        \item \textbf{TEE Attestation:} Verifies that the agent workload executed inside a trusted hardware enclave and produces an attestation report as a part of PoTE \cite{ref19, ref20}.
        \item \textbf{Oracle Attestation:} For off-chain verification of proof artifacts required in the Agent Contract. A quorum of oracles reach consensus on the proof object and set the report to validators on-chain for adding it to PoTE.
        \item \textbf{A-JWT ChecksumHash:} Issues a signed execution token (A-JWT) that includes deterministic hashes of the inputs/outputs and key execution metadata, enabling integrity checks and tamper-evident PoTE binding.
        \item \textbf{TLSNotary:} Generates cryptographic proof of specific HTTPS interactions (request/response commitments) to evidence off-chain actions performed by the agent \cite{ref18}.
        \item \textbf{Audit Rails:} Maintains a receipt ledger that records the full payment lifecycle as immutable events, so payment session can be audited end-to-end.
        \item \textbf{Telemetry Logs:} Captures structured runtime events (tool calls, model calls, policy checks, timestamps, errors) that reconstruct the execution trace and support verification, reconciliation, and review.
    \end{itemize}

\end{enumerate}

\subsubsection{Payment Settlement Plane}
\begin{list}{}{
    \setlength{\leftmargin}{2em} 
    \setlength{\parsep}{0pt}
}
\item \noindent This is the core payment settlement layer (see \textit{Fig. 1}) designed to be chain-agnostic. We identify the following key components:
\end{list}

\begin{itemize}
    \item \textbf{Rail Adapter:} A modular adapter that abstracts chain-specific transaction construction, gas/fee handling, confirmations, and finality into a single settlement interface. It lets TessPay plug in new chains or rails without changing the rest of the payment flow, while enforcing consistent status events and receipt formats across integrations.
    \item \textbf{Escrow Smart Contract:} A standardized escrow contract deployed per supported chain that locks funds, records settlement conditions, and releases value only when the required PoTE is submitted. It provides deterministic settlement outcomes (release or refund) and emits on-chain events that become the source of truth for completion.
    \item \textbf{Wallet Infrastructure:} A wallet stack for users and agents that manages keys, signing, and policy-based approvals across chains serving as an account manager \cite{ref22}. It funds escrow deposits, co-signs settlement actions, and supports secure key custody patterns while keeping settlement traceable to a wallet identity.
\end{itemize}

\subsection{Explore the Process Flows}
\setcounter{paragraph}{0}

\subsubsection{Orchestration Flow} 

\begin{itemize}
    \item \textbf{Task Submission:} As shown in \textit{Fig.~1, Step~1: Task Submission (indicated by blue)}, the user or merchant submits a task request to User Agent with required constraints. TessPay assigns canonical identifiers (\texttt{session\_id}, \texttt{workflow\_id} and \texttt{escrow\_id}) and initializes the control and verification plane workflow state. An initial ``task created'' receipt is emitted for auditability.

    \item \textbf{Mandate Generation:} User Agent converts the submitted intent into mandates that anchor the trust delegation involved in the agentic payment flow, \textit{see Fig.~1, Step~2: Mandate Generation (indicated by blue)}. This includes an intent mandate (capturing user intent), cart mandate (service cart), and payment mandate (payment authorisation). These mandates are stored as audit rails and hashed on-chain for immutability.

    \item \textbf{Identity Check and Authorization:} This step first performs an identity lookup by querying the agent registry to resolve the involved service agents to verifiable identities, ownership, and declared capabilities, and to fetch the metadata required for orchestration, \textit{see Fig.~1, Step~3: Identity Check and Authorization (in blue)}. Meanwhile the Identity and Authorization service also performs user-intent authorization by minting Agent JWT token embedding the authorized intent constraints.

    \item \textbf{Agent Retrieval:} Given the authorized task, the Orchestration Service queries the Off-chain Registry to discover eligible agents based on declared capabilities and availability. Candidate agents are ranked and routed based on their capabilities and similarity with the given task in the Orchestration Service (see Ranker and Router), \textit{see Fig.~1, Step~4: Agent Retrieval (indicated by blue)}.

    \item \textbf{Agent Assignment:} Orchestration Service chooses the service agent(s) and binds the assignment to \texttt{task\_id}, \textit{see Fig.~2, Step~5: Agent Assignment (indicated by blue)}. It sends an execution envelope to the Executor component for initialisation of the task delegation process. Meanwhile, the escrow contract has already held funds escrow (\textit{see Fig.~1, Escrow Smart Contract, Payment Settlement Plane}) awaiting PoTE.

    \item \textbf{Task Delegation:} The assigned service agent receives the execution envelope from the executor which includes the scope defined by user intent and begins execution, emitting telemetry and required Notary proofs as it progresses, \textit{see Fig.~1, Step~6: Task Delegation (in blue)}. Note that, the delegation by Executor is also captured as a TLSNotary proof. 
\end{itemize}

\subsubsection{Verification Flow}

\begin{itemize}
    \item \textbf{Output Submission:} The service agent submits the task output along with structured execution metadata bound to \texttt{workflow\_id} (and the related \texttt{escrow\_id}) to the facilitator, \textit{see Fig.~1, Step~1: Output Submission (in red)}. The submission triggers the verification pipeline and opens a verification session for receipt collection. 

    \item \textbf{TLS Receipts Generation:} TLSNotary receipts that cryptographically commit to agent actions including executor call, model call and tool call are submitted to the verification and audit service for verification, \textit{see Fig.~1, Step~2: TLS Receipts Generation (indicated by red)}. These receipts are linked to the execution artifacts via canonical identifiers so they can be independently validated.

    \item \textbf{Task and TLS Notary Verification:} These receipts are sent on-chain for verification where validators reach consensus following Byzantine Fault Tolerance (BFT) consensus algorithm \cite{ref21}. Once verified, these receipts along with Agent-JWT verification and TEE attestations are used to produce a Proof of Task Execution for this process, \textit{see Fig.~1, Step~3: Task and TLS Notary Verification (in red)}.

    \item \textbf{PoTE Submission:} The Verification service returns the finalized Proof of Task Execution bundle, composed of verified TLSNotary receipts, verified Agent-JWT and TEE attestations serving as a unified Proof of Task Execution \cite{ref26, ref27}. This PoTE is anchored to immutable audit events by storing the merkle root hash on-chain \cite{ref24}, \textit{see Fig.~1, Step~4: PoTE Submission (in red)}.

    \item \textbf{Performance Evaluation:} The Executor in Orchestration Service computes performance signals against predefined criteria (latency, success rate, constraint adherence, cost, and quality metrics) derived from telemetry, \textit{see Fig.~1, Step~5: Performance Evaluation (in red)}. 

    \item \textbf{Performance Disclosure:} The telemetry data captured is stored in audit rails and are available for future auditability, \textit{see Fig.~1, Step~6: Performance Disclosure (in red)}.
\end{itemize}

\begin{figure*}[htbp]
    \centering
    \includegraphics[width=\textwidth]{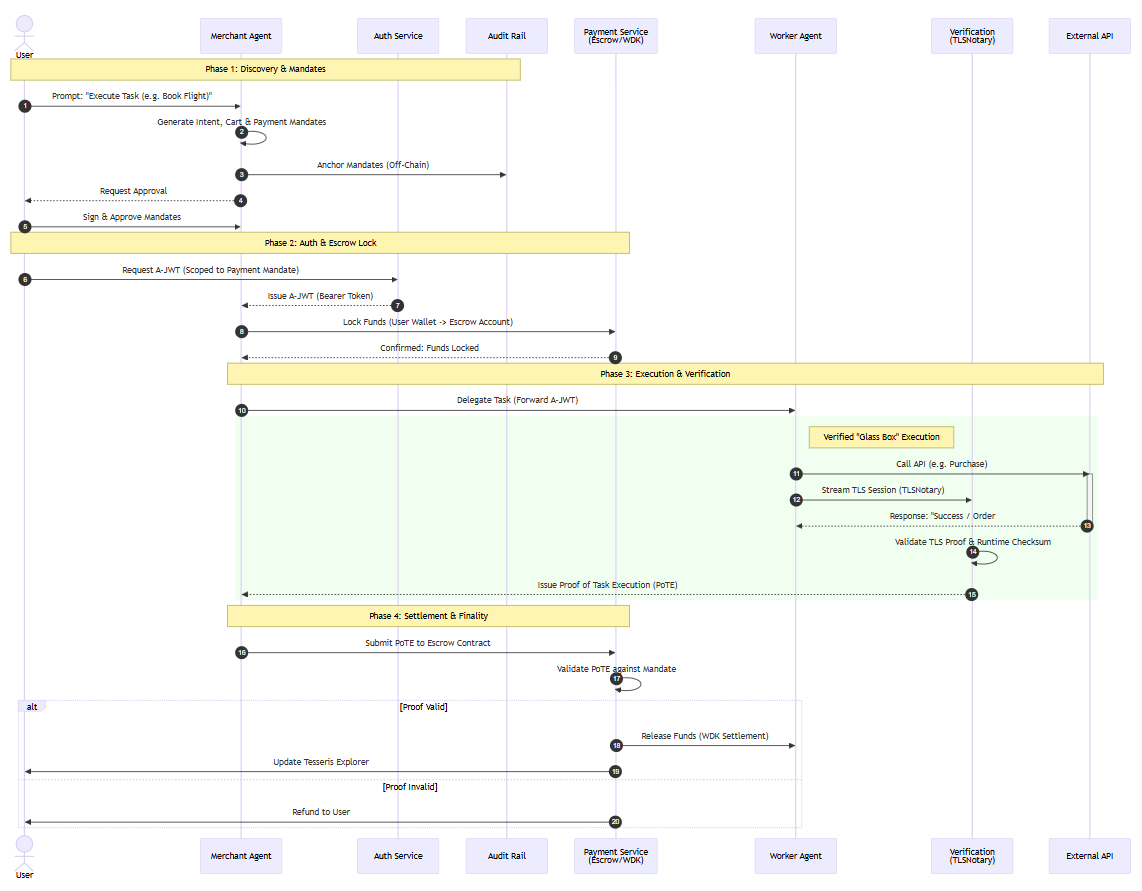} 
    \caption{\textbf{TessPay Payment Flow:} TessPay end-to-end transaction lifecycle across four phases - (1) discovery and mandate creation (Intent/Cart/Payment), anchored off-chain and user-approved; (2) scoped authorization (A-JWT) and escrow fund lock; (3) delegated task execution with “glass-box” verification via TLSNotary receipts and runtime checksums; and (4) PoTE-gated settlement, releasing funds to the worker agent on valid proof or refunding the user on failure.}
    \label{fig2}
\end{figure*}
\subsubsection{Payment Settlement Flow}

\begin{itemize}
    \item \textbf{Transfer to Escrow:} Funds are locked in an escrow smart contract associated with \texttt{escrow\_id}, deployed on the blockchain on which settlement is occurring. This transfer is bound to the payment mandate, \textit{see Fig.~1, Step~1: Transfer to Escrow (indicated in green)}. The release condition involves submission of PoTE serving as proof that the task was successfully executed. 

    \item \textbf{Payment Authorization:} Payment authorization is triggered only when the facilitator submits the PoTE Merkle root hash bound to the respective \texttt{workflow\_id} and texttt{escrow\_id}. TessPay verifies the root against the accepted proof set requirements, \textit{see Fig.~1, Step~2: Payment Authorization (in green)}.

    \item \textbf{Complete Escrow to Agent:} Upon successful verification of the merkle root, the facilitator makes a call to the escrow contract upon which it releases funds to the agent wallet as a part of its internal transaction, \textit{see Fig.~1, Step~3: Complete Escrow to Agent (in green)}.

    \item \textbf{Transaction Settlement:} The transaction is settled on the blockchain on which settlement is occurring. The PoTE merkle root hash and metadata are all captured and stored in the audit rail, \textit{see Fig.~1, Step~4: Transaction Settlement (in green)}.
\end{itemize}

\section{TessPay payment flow}
The TessPay payment flow occurs in the following phases:

\subsection{Discovery and Mandate Formulation}
This phase encapsulates the entire negotiation process. It begins with the Intent Mandate capturing the user's request. The Facilitator Agent creates a Cart Mandate after querying the worker. Crucially, once the user approves the cart, the Payment Mandate is immediately generated to lock in the financial terms. 

All three mandates are then indexed in the Audit Rail (Off-Chain DB):
\begin{enumerate}[label=\roman*.]
    \item \textbf{Intent Mandate:} Capturing the user's high-level request.
    \item \textbf{Cart Mandate:} Defining the specific items and pricing.
    \item \textbf{Payment Mandate:} Locking in the financial terms and authorization scope.
\end{enumerate}
This ensures that a complete, auditable contract exists before any sensitive authorization or fund movement occurs.

\subsection{Authorization and Escrow Locking}

With the Payment Mandate already stored and referenced, the Authorization Phase begins. The User Agent requests an Agentic JWT (A-JWT) from the IDP. The IDP validates the request against the pre-existing mandates to ensure the scope (\texttt{payment:escrow}) matches the user's locked agreement. 

Armed with this ``A-JWT Intent Token,'' the User Agent accesses the Credentials Database to fetch wallet keys and authorizes the Payment Service to move funds into Escrow.

\subsection{Verification}

Upon task delegation, the infrastructure enters a ``glass box'' execution mode where the Service Agent’s actions are simultaneously witnessed and measured. The Verification Service employs the TLSNotary protocol to cryptographically verify three distinct layers of interaction: the initial delegation from the Orchestration Executor, the reasoning inference from the AI Model, and the final tool execution against the External API. 

Running in parallel, the Telemetry Service captures granular runtime metrics---such as latency, throughput and token usage---storing the raw data in an off-chain database while anchoring a cryptographic hash of the session on-chain to ensure the integrity of the performance record.

Once the task is complete, the service aggregates these TLS proofs into a unified Proof of Task Execution (PoTE) document. Crucially, this document embeds the A-JWT Integrity Hash to cryptographically bind the proven actions to the specific identity and code version of the executing agent. The final PoTE is sealed by calculating its Merkle Root, which is then anchored on-chain. This creates an immutable, verifiable reference that allows the Escrow Smart Contract to validate the off-chain work before authorizing the release of funds.

\subsection{Settlement}

The settlement process begins when the Facilitator Agent submits the Proof of Task Execution (PoTE) to the Escrow Smart Contract. Acting as an autonomous financial judge, the contract cryptographically validates the PoTE’s signature and checks the on-chain Merkle Root to confirm the task's completion. Upon successful validation, the contract triggers the atomic release of funds, transferring the locked tokens from the Escrow Wallet directly to the Agent Wallet. This mechanism ensures that the financial outcome is inextricably linked to the verified execution state, preventing any premature or unauthorized disbursement of capital. By enforcing this conditionality at the bytecode level, TessPay eliminates the need for manual oversight during the final value transfer.

Once the on-chain transfer is complete, the Facilitator Smart Contract initiates the final verification loop. Functioning as a settlement oracle, it queries the underlying blockchain’s explorer to retrieve the transaction hash and validates that the \texttt{From}, \texttt{To}, and \texttt{Amount} fields match the original Payment Mandate. This secondary check serves as a reconciliation layer, confirming that the network-level execution perfectly mirrors the user's initial authorization. After confirming finality, the Facilitator indexes this event into the Tesseris Explorer, providing a unified, human-readable dashboard that displays the completed transaction status alongside the specific blockchain (e.g., Ethereum, Solana) where the settlement occurred. This indexing phase transforms raw ledger data into a structured audit trail, closing the lifecycle of the agentic payment.

\section{Integration of TessPay Across Chains}
TessPay’s cross-chain capabilities emerge directly from its Two-Plane Architecture as seen in Section~IV, which enforces a strict separation between the logic of verification and the mechanics of value transfer. The Control and Verification Plane governs intent capture, authorization, and state transitions, while the Settlement Plane is realized as a modular family of ``Rail Adapters'' that execute transfers through a unified settlement layer, enabling them to function independently.
\renewcommand{\thesubsection}{\Alph{subsection}} 
\setcounter{subsection}{0}

\subsection{Control and Verification Plane}
On the control side, TessPay composes the core microservices into a linear, rail-agnostic pipeline. This plane operates exclusively on abstract identifiers (e.g., \texttt{rail\_id}, \texttt{escrow\_id}, \texttt{task\_id}) rather than chain-specific primitives. The interaction between User Agent and Facilitator is captured in the form of verifiable credentials—Intent, Cart, and Payment Mandates—which are persisted by the Verification and Audit Service as Audit Rails and anchored as a hash on TessChain for future auditability. The Identity and Authorization Service generates an A-JWT token to ensure proper authorization at every workflow step. As the Service Agent executes, the Verification and Audit Service aggregates cryptographic proofs—TLSNotary receipts, verified A-JWT, and TEE Attestations—into a Proof of Task Execution (PoTE).

\subsection{Settlement Plane: Rail Adapters and Wallet Abstraction}
The Settlement Plane handles the concrete mechanics of asset movement. It standardizes interaction through a Rail-Generic Wallet Abstraction. Each supported blockchain is integrated via a dedicated Rail Adapter which includes a dedicated Escrow Smart Contract on that chain handling nuances such as gas fee estimation and finality confirmation. The plane exposes a minimal set of lifecycle operations to the Control Plane ensuring independent yet co-ordinated functioning between the two while also ensuring robust security.

\subsection{End-to-End Cross-Chain Flow}
By unifying these planes, the system achieves a consistent transactional flow across any supported rail:
\begin{itemize}
    \item \textbf{Rail Selection:} During mandate generation, a specific rail\_id is selected, binding the authorization scope to a concrete network.
    \item \textbf{Deposit \& Locking:} The User Agent triggers the Payment Service, causing the corresponding Rail Adapter to provision an escrow address. The adapter monitors this address until rail-specific finality is met, at which point the Control Plane transitions the escrow state to open.
    \item \textbf{PoTE-Gated Settlement:} Settlement is strictly gated by verification. Only when the Verification Service anchors a valid PoTE root does the Control Plane transition the escrow to settlement pending. This signals the Rail Adapter to sign and broadcast the payout transaction.
    \item \textbf{Refunds:} In cases of verification failure or timeout, the Control Plane instructs the adapter to execute a compensating refund transfer, ensuring atomic fairness.
\end{itemize}

\subsection{Extensibility}
This architecture ensures that onboarding a new settlement rail is a purely additive process. Integrating a new Layer-1 requires only the implementation of a Rail Adapter against the standard wallet interface and the registration of a new rail\_id along with deploying an Escrow smart contract on the chain. The core logic of the Control Plane---including mandate structures, A-JWT scopes, and PoTE verification---remains entirely invariant, allowing the ecosystem to scale horizontally across heterogeneous blockchain networks with zero friction.


\section{ Tiered Service Model}
Recognizing that not all agentic transactions carry the same risk profile, TessPay implements a Tiered Service Model. The Control Plane (TessChain) dynamically applies different levels of verification rigor and settlement latency based on the transaction volume and value defined in the Payment Mandate. This ``Risk-Adaptive'' approach ensures that micro-transactions remain efficient and low-cost, while high-value transfers are subjected to maximum cryptographic scrutiny.

\subsection*{Tier 1: Low-Risk Transactions}

\subsubsection*{A. Target Profile} High-frequency, low-value interactions (e.g., streaming API payments, content access fees, or micro-tipping under \$10).

\subsubsection*{B. Verification Logic: Optimistic Execution}
\begin{itemize}

\item \textbf{Mechanism:} For these transactions, the overhead of a full TLSNotary session for every single cent is inefficient. Instead, the system relies primarily on the A-JWT for authorization.
\item \textbf{Settlement:} The Verification Service logs the activity but may bypass the full MPC-TLS witness process in favor of a simpler ``API Receipt'' check. The Payment Service aggregates these micro-charges into a single ``Batch Settlement'' on the Settlement Plane to minimize gas fees.
\item \textbf{Security Trade-off:} The system accepts a marginal risk of dispute in exchange for sub-second latency and minimal cost.
\end{itemize}

\subsection*{Tier 2: Medium-Risk Transactions}

\subsubsection*{A. Target Profile}  The standard operational tier for daily economic activity (e.g., e-commerce purchases, utility bill payments, or standard token swaps ranging from \$10 to \$1{,}000).
\subsubsection*{B. Verification Logic: The Standard PoTE Flow}
\begin{itemize}
\item \textbf{Mechanism:} This tier employs the full architectural suite described in Section~4. Every transaction requires a dedicated TLSNotary session to witness the execution.
\item \textbf{Settlement:} Funds are locked in individual Escrow Accounts. The release is strictly gated by the generation of a complete Proof of Task Execution (PoTE) containing the A-JWT Integrity Hash and the TLS Proofs.
\item \textbf{Security Trade-off:} This tier represents the optimal balance, offering cryptographic finality and non-repudiation with a processing time suitable for standard commerce (seconds to minutes).
\end{itemize} 

\subsection*{Tier 3: High-Risk Transactions}
\subsubsection*{A. Target Profile}  
Large-scale asset management, B2B procurement, or Whale portfolio rebalancing (transactions exceeding \$1{,}000 or involving sensitive assets).
\subsubsection*{B. Verification Logic: Defense-in-Depth Verification} 
\begin{itemize}
\item\textbf{Mechanism:} For these critical transactions, the Verification Service enforces Multi-Witness Validation. A single TLSNotary proof is insufficient; the system requires corroboration from multiple decentralized notaries or additional TEE (Trusted Execution Environment) Attestation (certifying the agent’s hardware isolation).
\item\textbf{Settlement:} The Settlement Plane enforces an ``Extended Finality'' period. The Rail Adapter waits for a higher number of block confirmations on the underlying chain before confirming the transaction to the Control Plane.
\item\textbf{Security Trade-off:} Latency and cost are secondary to absolute security. The system introduces deliberate friction---requiring multi-signature mandates or a ``Challenge Period''---to ensure that any anomaly can be intercepted before irreversible settlement occurs.
\end{itemize}



\section{Security and Threat Modelling}
To evaluate the robustness of the TessPay Multi-Rail Architecture, we analyze the system against specific high-risk attack vectors. We employ a scenario-based methodology, focusing on threats to fund integrity, execution verification, and key management \cite{ref15, ref17}. The analysis demonstrates how the architectural decoupling of the Settlement Plane and the Control Plane (TessChain) provides intrinsic mitigation against these threats.

\subsection*{Scenario A: The Phantom Deposit (Oracle Manipulation)}

\subsubsection*{A. Attack Vector}
A malicious user attempts to initiate a task and trigger an OPEN escrow state on TessChain without actually locking funds on the external settlement rail (e.g., by broadcasting a transaction that immediately reverts or by providing a fraudulent transaction hash).

\subsubsection*{B. Mitigation Strategy}
The system mitigates this through the strict Deposit Observation workflow inherent to the Settlement Plane. TessChain does not accept user-submitted proofs of deposit blindly. Instead, the Rail Adapter independently observes the blockchain state via its configured JSON-RPC endpoints. Furthermore, the Rail Adapter is configured to recognize an inbound transfer to the escrow account only after a rail-specific finality threshold is met (e.g., block confirmations on EVM).

\subsubsection*{C. Resolution}
An EscrowRecord is only instantiated or transitioned to OPEN status when the trusted Rail Adapter confirms the validity of the \texttt{deposit\_tx\_id} and the amount via the explorer. This ensures that the Control Plane state remains synchronized with the ground truth of the Settlement Plane.

\subsection*{Scenario B: The Unverified Payout (Bypassing PoTE)}

\subsubsection*{A. Attack Vector}
A compromised or malicious agent attempts to force a \texttt{settle()} call on the Rail Adapter to withdraw funds from the escrow account without performing the assigned task or providing a valid Proof of Task Execution (PoTE).

\subsubsection*{B. Mitigation Strategy}
This threat is neutralized by the PoTE-Gated State Machine enforced by the Control and Verification Plane. The Rail Adapter, which holds the keys to move funds, effectively operates as a ``dumb'' executor; it exposes the \texttt{settle()} API but lacks the autonomous logic to invoke it. The logic to trigger settlement resides exclusively on TessChain, requiring the PoTE subsystem to verify and anchor a \texttt{pote\_hash}.

\subsubsection*{C. Resolution}
Without a valid PoTE hash linked to the specific \texttt{task\_id} in the TessChain state, the instruction to call \texttt{settle()} is never generated. Consequently, the agent never receives a signing request for the payout, rendering the funds immobile even if the agent interacts directly with the Control Plane APIs.

\subsection*{Scenario C: Control Plane Compromise (Key Exfiltration)}
\subsubsection*{A. Attack Vector}
An attacker gains administrative access to the TessChain nodes or the TessIndex database, attempting to exfiltrate private keys to drain user and escrow funds across multiple rails.

\subsubsection*{B. Mitigation Strategy}
The architecture’s Two-Plane Model provides a robust defense-in-depth mechanism through the Principle of Least Privilege. TessChain acts solely as a registry of signed facts (amounts, addresses, proofs). It does not generate, store, or manage private keys. All key management and signing operations are encapsulated within the wallet infrastructure residing in the Settlement Plane.

\subsubsection*{C. Resolution}
Even in the catastrophic event of a total Control Plane compromise, the attacker obtains only metadata (escrow IDs and status logs). They cannot construct valid raw transactions on the external rails because the cryptographic material remains isolated within the Rail Adapter’s secure storage environment.

\subsection*{Scenario D: Cross-Rail Replay and Confusion}
\subsubsection*{A. Attack Vector}
In a multi-rail environment, an attacker attempts to replay a settlement signature from a testnet rail (e.g., Sepolia) onto a mainnet rail (e.g., Ethereum Mainnet) to drain real funds using testnet credentials.

\subsubsection*{B. Mitigation Strategy}
The system prevents this through strict Rail Abstraction and Typing. The \texttt{WalletModule} interface requires a strict \texttt{rail\_id} (e.g., \texttt{evm:sepolia}). The Rail Adapter binds specific WDK instances to specific chain configurations, while the \texttt{AgentWallet} schema binds an identity to a specific \texttt{rail\_id} and address.

\subsubsection*{C. Resolution}
A signed transaction generated for \texttt{evm:sepolia} includes chain-specific replay protection (such as EIP-155 Chain ID enforcement). Furthermore, the TessChain EscrowRecord strictly scopes \texttt{settlement\_tx\_ids} to the originating \texttt{rail\_id}, causing any chain-mismatched hashes to be rejected by the observation logic.

\section{Use Case}

\subsection{E-Commerce Shopping Agent}

Consider this scenario---a user searching for ``durable noise-canceling headphones'' has to open 10+ tabs, read conflicting Reddit reviews, compare technical specs (drivers, battery life), and check for cross-site price variations. A shopping agent solves this problem easily with multi-dimensional filtering, processing thousands of data points in seconds to find the optimal product based on user intent. 

Beyond this, users typically have to search for products manually based on their own research (e.g., ``Hyaluronic Acid'' vs. ``Salicylic Acid'') rather than searching directly based on the desired outcome (e.g., ``I need a skincare routine for dry skin under \$50''). An e-commerce agent solves this by mapping the user's goal to specific product specifications.

But the most important barriers to agentic commerce in this scenario are ``trust'' and ``security.'' For an agent to make a purchase, it requires user credentials like shipping addresses and credit card details---a massive security risk. Furthermore, there are concerns regarding agents hallucinating and buying incorrect items, alongside a lack of infrastructure to define liability and effective dispute resolution. Finally, an agent is always susceptible to injection attacks, undermining trust. The TessPay infrastructure overcomes these barriers effectively:

\subsubsection{User Credentials}
User credentials are stored in a separate database from the user agent and can only be accessed by an Agent-JWT (A-JWT) token authorized by the specific user's intent. This token is minted by the Identity and Authorization service by explicitly capturing the user's authorization.

\subsubsection{Agent Hallucinations and Dispute Resolution}
Agent hallucinations are addressed by grounding agent actions in verifiable credentials called mandates. The user's intent is captured as an \textit{intent mandate} guiding the agent's action, and a \textit{cart mandate} captures the output returned by the agent in the form of a checkout cart. These mandates define accountability and are stored in an off-chain database as \textit{Audit Rails} for future dispute resolution.

\subsubsection{Injection Attacks}
The agent is executed within a Trusted Execution Environment (TEE) container as a defense against injection attacks. During the verification process, the \textit{Proof of Task Execution (PoTE)} includes a TEE Attestation as cryptographic proof that the agent remained within the TEE during the entire payment flow.

\subsection{Portfolio Manager Agent}

Consider a user who desires the following portfolio: ``medium-risk, diversified, monthly rebalance, no leverage, and strict risk limits.'' Current solutions involve manually aggregating holdings across brokers, exchanges, and wallets; calculating drift vs. target allocation; screening assets; estimating slippage; and placing orders. A Portfolio Manager Agent automates this process by converting goals into constraints, generating a trade plan using sophisticated algorithms, and executing trades across venues.

Again, the key barriers are trust and security. Executing trades requires access to high-risk credentials (API keys, wallet signing), which expands the blast radius of any compromise. Agents can also make financially destructive mistakes (wrong instrument mapping, sizing errors, or violating policy). Without structured evidence, it is hard to prove what the user authorized versus what the agent executed. This is compounded by adversarial-input risk, where manipulated news or research could trigger unauthorized trades.

\subsubsection{User Credentials}
TessPay keeps user credentials in a separate store from the agent and only releases scoped access through an A-JWT minted by the Identity and Authorization service. The token is bound to the specific portfolio intent, ensuring the agent cannot exceed the user's approved scope.

\subsubsection{Hallucinations and Dispute Resolution}
TessPay grounds actions in verifiable mandates:
\begin{enumerate}[label=\roman*.]
    \item \textbf{Intent Mandate:} Records user objectives and risk posture.
    \item \textbf{Cart Mandate:} Formalizes constraints and the trade plan.
    \item \textbf{Execution Receipt Mandate:} Records fills, fees, and transaction hashes.
    \item \textbf{Portfolio Snapshot Mandate:} Records before/after states.
\end{enumerate}
These are stored as Audit Rails to support deterministic accountability.

\subsubsection{Verify-then-Pay}
TessPay releases the user-authorized fee only after the \textit{PoTE} is verified. This ensures the agent's trade plan complied with policy and that execution receipts match the approved plan within defined tolerances. Payment is conditional on measurable compliance rather than subjective claims.

\subsubsection{Injection Attacks}
Running the agent inside a TEE and including a TEE attestation in the PoTE cryptographically proves the agent operated in a hardened runtime. This reduces tampering risk and makes injection-driven deviations materially harder to execute without detection.

\section{Future Work}

\subsection{Improving PoTE Robustness and Verifiability}

A primary focus for future development is enhancing the resilience of our consensus model by evolving it into a Universal Cross-Chain Verification Layer. By extending the validator set to operate across heterogeneous ecosystems, we mitigate the risks of siloed finality and localized ledger attacks. This establishes a ubiquitous "Layer 0" for agentic trust, where a single, decentralized quorum secures transactions across fragmented economies, ensuring that verification remains robust even if underlying settlement rails experience volatility or isolation.

In parallel, we plan to address the inherent non-determinism of LLMs by integrating probabilistic models for semantic verification to assess the contextual quality of agent outputs. By evaluating the semantic alignment between the user's original mandate and the final execution artifacts, the system can detect subtle deviations or partial failures that strictly binary proofs might miss. This ensures that payment release is conditional on both logic-integrity and qualitative-utility, confirming that tasks fully satisfy the user's nuanced intent while protecting against "lazy" or "hallucinated" outputs that might bypass standard technical checks.

\subsection{Advanced Threat Modeling and Adaptive Security}
As autonomous agent ecosystems evolve, so do the attack vectors targeting them. Future work will concentrate on dynamic threat detection and the hardening of the User Authorization Service. We intend to iterate on our threat model to specifically address ``Excessive Agency'' and sophisticated \textit{Prompt Injection} attacks that may bypass static checksums. 

This includes researching adaptive security policies that can analyze agent behavior in real-time, automatically revoking A-JWT permissions if anomalous execution patterns or unauthorized delegation chains are detected. Furthermore, we plan to formalize the standardization of the ``Agentic JWT'' protocol within the wider security community, subjecting the \texttt{agent\_checksum} and \texttt{delegation\_chain} claims to broader peer review to ensure they meet rigorous industry standards for \textit{Zero Trust} architectures.

\subsection{Seamless Web2 and Web3 Interoperability}
While the current TessPay architecture is optimized for decentralized, chain-agnostic settlements, a critical next step is bridging the gap between Web3 financial rails and Web2 legacy infrastructure. Future research will focus on extending the Mandate system to support hybrid payment rails, allowing agents to seamlessly switch between cryptocurrency escrow and traditional fiat payment gateways (e.g., Stripe, PayPal) within a single workflow. 

This interoperability initiative aims to create a universal ``Agentic Payment Standard'' where an A-JWT can authorize a transaction regardless of whether the settlement occurs on a blockchain ledger or a centralized banking network. By abstracting the settlement layer further, we aim to enable agents to operate fluidly across both domains, reducing the friction for enterprise adoption of agentic commerce.

\section{Conclusion}
This paper addressed a critical trust gap in agentic commerce: the mismatch between infrastructure designed for human interaction and the requirements of autonomous operation by agents. We analyzed the missing primitives across the transaction lifecycle, from task delegation which lacks mechanisms to define scope and authorize actions based on user intent to verified settlement, where payments settle optimistically without proof of execution, and dispute resolution, which suffers from lack of audit mechanisms for providing accountability in case of disputes and refunds. Our review of existing protocols, including Google’s AP2, Coinbase’s x402, ACP, and ERC-8004, reveals a fragmented landscape where no single solution addresses the end-to-end trust gap. Current approaches largely rely on optimistic 'pay-then-consume' models or reputation scoring, while others tightly couple reasoning with key custody, exposing users to prompt injection risks and unauthorized fund access. These limitations underscore the critical need for a unified infrastructure that binds delegation, secure execution, and verifiable settlement into a single, auditable lifecycle.

To close this gap, we introduced TessPay, a unified infrastructure for verified agentic payments built on a "Verify-then-Pay" settlement model. We outlined a dual-plane architecture comprising a Control and Verification Plane and a Payment Settlement Plane, detailing the microservices that constitute each. The Control and Verification Plane includes an Identity and Authorization Service to capture user intent and authorize actions, an Orchestration Service to manage the end-to-end workflow, and a Verification Service to validate cryptographic evidence of task execution. Complementing this, the Payment Settlement Plane leverages modular payment rails, supported by an Escrow and Wallet Infrastructure that enforces conditional, verified settlement. Finally, we examined the internal process flows for each function and presented a unified, end-to-end payment lifecycle that integrates these components into a seamless transaction.

At the workflow level, TessPay operationalizes trust across four key stages. Before execution, a canonical registry anchors agents with stable identities to enable intent-driven discovery, while authority is delegated through verifiable mandates that ensure stakeholder accountability. During execution, funds are secured in escrow, awaiting validation, while the Verification Service captures execution receipts as cryptographic evidence. At settlement, the system validates these receipts to generate the Proof of Task Execution (PoTE), triggering the escrow to release funds only upon successful verification. After settlement, TessPay preserves a tamper-evident audit trail, ensuring clear accountability for dispute resolution and refunds.

Next, we explored how TessPay enables chain-agnostic settlement across heterogeneous chains. The architectural separation of control from settlement allows for the parallel integration of modular payment rails within the Settlement Plane. We then defined TessPay’s tiered service model, which handles transactions based on their risk profile, assigning robust verification mechanisms (TLSNotary or TEE attestations) proportionate to the value and sensitivity of the exchange.

We evaluated TessPay's security via scenario-driven threat modeling, demonstrating how the two-plane architecture and microservices within mitigate key failure modes, including phantom deposits, PoTE bypass attempts, control-plane compromise, and cross-rail replay attacks. To demonstrate practical applicability, we presented two representative use cases—an e-commerce shopping agent and a portfolio manager agent—illustrating how the TessPay infrastructure resolves critical bottlenecks of trust and security in real-world scenarios. Finally, we outlined directions for future work, focusing on improving the robustness of PoTE verifiability, enhancing security considerations, and enabling seamless Web2 and Web3 interoperability to move towards a universal agentic payment standard.

\end{document}